\documentclass[journal]{IEEEtran}
\usepackage{cite}
\usepackage{amsfonts,amssymb}
\usepackage{amsbsy}
\usepackage{graphicx}
\usepackage{latexsym}
\usepackage{amsmath,mathrsfs}
\usepackage{psfrag}
\usepackage{subfigure}
\newtheorem{Def}{Definition}
\newtheorem{Eg}{Example}
\newtheorem{Prop}{Proposition}
\newtheorem{Thm}{Theorem}
\newtheorem{Lem}{Lemma}
\newtheorem{Coro}{Corollary}
\newtheorem{Rem}{Remark}

\newcommand{\A}{\mathcal{A}}
\newcommand{\B}{\mathcal{B}}
\newcommand{\E}{\mathcal{E}}
\newcommand{\F}{\mathcal{F}}
\newcommand{\K}{\mathcal{K}}
\newcommand{\M}{\mathcal{M}}
\newcommand{\LL}{\mathcal{L}}
\newcommand{\OO}{\mathcal{O}}

\begin{document}
\title{Supervisory Control of Fuzzy Discrete Event Systems}
\author{Yongzhi Cao and Mingsheng Ying%
\thanks{This work was supported by the National Foundation of Natural Sciences of China (Grant No: 60273003) and the
Key Grant Project of Chinese Ministry of Education.}%
\thanks{The authors are with the State Key Laboratory of Intelligent Technology and
Systems, Department of Computer Science and Technology, Tsinghua
University, Beijing 100084, China (e-mails:
caoyz@mail.tsinghua.edu.cn (Y.Z. Cao);
yingmsh@mail.tsinghua.edu.cn (M.S. Ying)).}}

\markboth{CAO AND YING: Supervisory Control of Fuzzy Discrete Event Systems}{CAO AND YING:
Supervisory Control of Fuzzy Discrete Event Systems}
\maketitle

\begin{abstract}
In order to cope with situations in which a plant's dynamics are not precisely known, we consider
the problem of supervisory control for a class of discrete event systems modelled by fuzzy
automata. The behavior of such discrete event systems is described by fuzzy languages; the
supervisors are event feedback and can disable only controllable events with any degree. The
concept of discrete event system controllability is thus extended by incorporating fuzziness. In
this new sense, we present a necessary and sufficient condition for a fuzzy language to be
controllable. We also study the supremal controllable fuzzy sublanguage and the infimal
controllable fuzzy superlanguage when a given pre-specified desired fuzzy language is
uncontrollable. Our framework generalizes that of Ramadge-Wonham and reduces to Ramadge-Wonham
framework when membership grades in all fuzzy languages must be either $0$ or $1$. The theoretical
development is accompanied by illustrative numerical examples.
\end{abstract}

\section{Introduction}
\PARstart{A}{discrete} event system (DES) is a dynamic system with discrete states that are
event-driven. The principal features of such systems are that they are discrete, asynchronous, and
possibly nondeterministic. DES arise in a variety of contexts ranging from computer operating
systems to the control of complex multimodel processes. DES theory, particularly on modelling and
control, has been successfully employed in many areas such as concurrent program semantics,
monitoring and control of complex systems, and communication protocols. The behavior of DES is
described as the sequences of occurrences of events involved. In order to restrict the behavior of
a DES to a desired range, supervisory control theory initiated by Ramadge and Wonham \cite{RW87}
has been intensively investigated over the last decades (see, for example, \cite{KG95,CL99}).

Usually, a DES is modelled by finite state automaton with events as input alphabets, and the
behavior is thus the language generated by the automaton. Most of the published research on DES has
been concentrated on systems that are modelled as deterministic automata or nondeterministic
automata with $\epsilon$-transitions. A common feature of these models is that their state
transitions are crisp once that corresponding events occur, that is, no uncertainty arises in the
state transitions. There are, however, situations in which the state transitions of some systems
are always somewhat imprecise, uncertain, and vague even in a deterministic sense. A typical
example given in \cite{LY01,LY02} is a person's health status, where the change of the condition of
a person's health from a state, say ``excellent", to another, say ``good", is obviously imprecise,
since it is hard to measure exactly the change. More examples will be encountered when we examine
chemical reactions, mobile robots in an unstructured environment \cite{MG02}, intelligent vehicle
control \cite{R03}, wastewater treatment \cite{WSEAD00}, and so on.

Uncertainty arises not only in state transitions, but also in states themselves and control
strategies. The notion of a fuzzy subset of a set introduced by Zadeh in 1965 \cite{Z65} has been
 well applied to a wide range of scientific areas and has been proved to be a good method for
representing uncertainty. To capture uncertainty appearing in states and state transitions of DES,
Lin and Ying incorporated fuzzy set theory together with DES and thus extended crisp DES to fuzzy
DES by proposing fuzzy finite automaton model in \cite{LY01,LY02}. Under the framework of fuzzy
DES, they studied generalized observability and some optimal control problems. Taking into account
the uncertainty associated with control strategies, Lavrov \cite{L03} introduced an enhancement of
industrial control by combining discrete event supervisory control methods and fuzzy techniques.

In recent years, fuzzy model such as fuzzy automata, fuzzy Petri nets, and fuzzy neural networks,
as a complement to conventional models, has become an active research topic and found successful
applications in many areas \cite{DDP04,GF02,PR03,R03,WSEAD00}. In terms of DES, fuzzy Petri net
model has been investigated for many years \cite{ZZ94},\cite{HLE01,RMZ03}; however, to our
knowledge, few efforts, except the work \cite{LY01,LY02}, are made to consider the model and
supervisory control following for fuzzy automata.

The purpose of this paper is to develop supervisory control theory for fuzzy DES by extending
traditional supervisory control theory for crisp DES. The generalized supervisory control theory
describes a class of DES that state transitions are imprecise and uncertain by (maxmin) fuzzy
automata introduced in \cite{S68}; the behavior of such systems is described by their generated
fuzzy languages. Control is exercised by a fuzzy supervisor that disables events with certain
degrees in the controlled system, the plant, so that the closed-loop system of supervisor and plant
exhibits a pre-specified desired fuzzy language. We will introduce the notion of (fuzzy)
controllability for fuzzy languages by generalizing crisp controllability, and present a necessary
and sufficient condition for a fuzzy supervisor synthesis. When a desirable fuzzy language is not
controllable, we will investigate its supremal controllable fuzzy sublanguage and infimal
controllable fuzzy superlanguage, and provide their expressions and algorithms. Our results contain
those relevantly known characterizations of crisp DES, as expected.

Several related researches should be distinguished before introducing the organization of the
paper. The work presented here is a continuation of \cite{LY01,LY02}, where the control problems
studied are based on states and state sequences. We are now concerned with event feedback control
with fuzzy supervisor; in addition, our fuzzy automaton model also differs from that of
\cite{LY01,LY02}. Our fuzzy supervisor is similar to the probabilistic supervisor introduced in
\cite{LW93}, but resultant plant dynamics are quite distinct from each other. Fuzzy languages are
similar to probabilistic languages \cite{P71},\cite{GKM99,KG01}, but their semantics are different:
the weight of a string in a probabilistic context reflects a frequency of occurrence, while the
weight in a fuzzy context describes the membership grade (namely, uncertainty) of a string. It is
well known that probability theory is not capable of capturing uncertainty in all its
manifestations.

The rest of the paper is organized as follows. In Section II, we sketch some basics of crisp DES
control and fuzzy set theory. In Section III, we introduce the base model of fuzzy DES. The concept
of generalized controllability and a controllability theorem for fuzzy languages are given in
Section IV. Section V and VI are devoted to investigating the supremal controllable fuzzy
sublanguage and the infimal controllable fuzzy superlanguage of an uncontrollable fuzzy language,
respectively. The paper is concluded in Section VII with a brief discussion on the future research.

\section{Preliminaries}
In the following two subsections, we will briefly recall the background on supervisory control of
crisp DES and a few basic facts on fuzzy set theory.
\subsection{Crisp Discrete Event Systems}
Let $E$ denote the finite set of events, and $E^*$ denote the set of all finite strings constructed
by concatenation of elements of $E$, including the empty string $\epsilon$. A string $\mu\in E^*$
is a prefix of a string $\omega\in E^*$ if there exists $\nu\in E^*$ such that $\mu\nu=\omega$. In
this case, we write $\mu\leq\omega$. For any $\omega\in E^*$ with length at least $1$, we use
$\tilde{\omega}$ to denote the maximal proper prefix of $\omega$ in what follows. In other words,
$\tilde{\omega}a=\omega$ for some $a\in E$. We write $|\omega|$ for the length of a string
$\omega$. Any subset of $E^*$ is called a language over $E$. The prefix closure of a language $L$
consists of the set of strings which are prefixes of strings in $L$. More precisely, the prefix
closure of $L$, denoted by $\overline{L}$, is defined by $\overline{L}=\{\mu\in E^*:\mu\leq\omega\
\textrm{for some}\ \omega\in L\}$. A language $L$ is said to be prefix closed if $L=\overline{L}$.

A DES, or plant, is usually described by a deterministic automaton: $G=(Q,E,\delta,q_0)$, where $Q$
is a set of states, with the initial state $q_0$, $E$ is a set of events, and $\delta:Q\times
E\rightarrow Q$ is a transition function (in general a partial function). The function $\delta$ is
extended to $\delta:Q\times E^*\rightarrow Q$ in the obvious way. In a logical model of a DES, we
are interested in the strings of events that the system can generate. Thus the behavior of a DES is
modelled as a prefix closed language $L(G)=\{s\in E^*:\delta(q_0,s)\mbox{ is defined}\}$ over the
event set $E$.

The supervisory control theory partitions the event set of the system into the two disjoint sets of
controllable and uncontrollable events, $E_c$ and $E_{uc}$, respectively. A supervisor is a map
$S:L(G)\rightarrow2^E$ such that $S(s)\supseteq E_{uc}$ for any string $s\in L(G)$. The language
generated by the controlled system is denoted by $L(S/G)$ which is defined inductively as follows:

1) $\epsilon\in L(S/G);$

2) $[(s\in L(S/G))$ and $(sa\in L(G))$ and $(a\in S(s))]\Leftrightarrow[sa\in L(S/G)].$

Let us review the controllability definition of (crisp) languages  due to Ramadge and Wonham in
\cite{RW87}.

\begin{Def}A language $K\subseteq L(G)$ is said
to be {\it controllable} (with respect to $L(G)$ and $E_{uc}$) if $$\overline{K}E_{uc}\cap
L(G)\subseteq\overline{K}.$$\label{Dcont}
\end{Def}

The key existence result for a supervisor is the following \cite{RW87}.
\begin{Prop}
Given a nonempty language $K\subseteq L(G)$, there exists a supervisor $S$ such that
$L(S/G)=\overline{K}$ if and only if $K$ is controllable with respect to $L(G)$ and $E_{uc}$.
\end{Prop}

A good property of controllable languages is that they are closed under arbitrary unions. Thus, for
any language $K$, there exists the supremal controllable sublanguage of $K$ \cite{WR87}, which is
obtained by taking the union of all controllable sublanguages of $K$. Another somewhat weak
property of controllable languages is that arbitrary intersections of prefix closed and
controllable languages are again prefix closed and controllable. As a result, the intersection of
all prefix closed and controllable suplanguages of $K$ gives rise to the infimal prefix closed
controllable suplanguages of $K$ \cite{LC90}.

\subsection{Fuzzy Set}
Assume now that $X$ is a universal set. Then, every subset $A$ of $X$ can be uniquely represented
by a so-called characteristic function $\chi_A:X\rightarrow\{0,1\}$, which is given by
\begin{displaymath}
\chi_A(x)=\left\{ \begin{array}{ll}
1, & \textrm{if $x\in A$}\\
0, & \textrm{if $x\not\in A$}
\end{array} \right.
\end{displaymath}for any $x\in X$. Thus the boundary of a set is required to be precise. That is, a
set is a collection of things for which it is known whether any given thing is inside it or not.
Contrary to classical crisp sets, fuzzy sets do not have sharp boundaries. That is, being a member
of a fuzzy set is not a simple mater of being definitely in or definitely out: A member may be
inside the set to a greater or lesser degree. For example, the set of tall people is a set whose
exact boundary cannot be precisely determined. Suppose that we define all tall people as those
having a height greater than or equal to $1.8$ meters. Then a person whose height is $1.79$ meters
will not be considered a tall person. However, according to our intuition, there is no clear
distinction between a person of $1.79$ meters and a person of $1.8$ meters in terms of the word
``tall". If we allow that the same person be considered ``tall" to some degree and be considered
``not tall" to another degree, we will not be in a dilemma.

Each {\it fuzzy subset} (or simply {\it fuzzy set}), $\A$, is defined in terms of a relevant
universal set $X$ by a function assigning to each element $x$ of $X$ a value $\A(x)$ in the closed
unit interval $[0,1]$. Such a function is called a membership function, which is a generalization
of the characteristic function mentioned above; the value $\A(x)$ characterizes the degree of
membership of $x$ in $\A$. Accordingly, a person is a member of the set ``tall people" to the
degree to which he or she meets the operating concept of ``tall".

The {\it support} of a fuzzy set $\A$ is a crisp set defined as $supp(\A)=\{x:\A(x)>0\}$. Whenever
$supp(\A)$ is a finite set, say $supp(\A)=\{x_1,x_2,\cdots,x_n\}$, then fuzzy set $\A$ can be
written in Zadeh's notation as
follows:$$\A=\frac{\A(x_1)}{x_1}+\frac{\A(x_2)}{x_2}+\cdots+\frac{\A(x_n)}{x_n}.$$

We denote by $\F(X)$ the set of all fuzzy subsets of $X$. For any $\A,\B\in \F(X)$, we say that
$\A$ is contained in $\B$ (or $\B$ contains $\A$), denoted by $\A\subseteq\B$, if $\A(x)\leq\B(x)$
for all $x\in X$. We say that $\A=\B$ if and only if $\A\subseteq\B$ and $\B\subseteq\A$.
Obviously, we see that $\A\subseteq\B$ implies $supp(\A)\subseteq supp(\B)$; the latter is the
usual set inclusion. A fuzzy set is said to be {\it empty} if its membership function is
identically zero on $X$. We use $\OO$ to denote the empty fuzzy set.

For any family $\lambda_i,\ i\in I$, of elements of $[0,1]$, we write $\vee_{i\in I}\lambda_i$ or
$\vee\{\lambda_i:i\in I\}$ for the supremum of $\{\lambda_i:i\in I\}$, and $\wedge_{i\in
I}\lambda_i$ or $\wedge\{\lambda_i:i\in I\}$ for its infimum. In particular, if $I$ is finite, then
$\vee_{i\in I}\lambda_i$ and $\wedge_{i\in I}\lambda_i$ are the greatest element and the least
element in $\{\lambda_i:i\in I\}$, respectively.

Let $\A,\B\in \F(X)$. The {\it union} of $\A$ and $\B$, denoted by $\A\cup\B$, is defined by the
membership function $$(\A\cup\B)(x)=\A(x)\vee\B(x)$$ for all $x\in X$; the {\it intersection} of
$\A$ and $\B$, denoted by $\A\cap\B$, is given by the membership function
$$(\A\cap\B)(x)=\A(x)\wedge\B(x)$$ for all $x\in X$.

\section{Formalism for Fuzzy Discrete Event Systems}In this section, we introduce fuzzy automaton model
and fuzzy languages generated by the model.

As was mentioned in the introduction, we would like to formalize a class of DES, where the state
sets are crisp and the state transitions are uncertain. We adopt a kind of fuzzy automata, which is
known as maxmin automata in some mathematical literature \cite{S68,KL79} and is somewhat different
from max-product automata used in \cite{LY01,LY02}, to model the systems. Formally, we have the
following.

\begin{Def}A {\it fuzzy automaton} is a four-tuple $G=(Q, E,
\delta, q_0)$, where:

$Q$ is a crisp (finite or infinite) set of states,

$E$ is a finite set of events,

$q_0\in Q$ is the initial state, and

$\delta$ is a function from $Q\times E\times Q$ to $[0,1]$, called a fuzzy transition function.
\end{Def}

For any $p,\ q\in Q$ and $a\in E$, we can interpret $\delta(p,a,q)$ as the possibility degree to
which the automaton in state $p$ and with the occurrence of event $a$ may enter state $q$. In our
graph representation of automata later, $\delta$ will be represented by arcs between states, and
labels on the arcs. If $p,q$ are states and $a$ is an event such that $\delta(p,a,q)>0$, then we
make an arc labelled $a|\delta(p,a,q)$ from $p$ to $q$.

In contrast with crisp DES, an event in fuzzy DES may take the system to more than one states with
different degrees. The concept of fuzzy automata is a natural generalization of nondeterministic
automata. The major difference between fuzzy automata and nondeterministic automata is: in a
nondeterministic automaton, $\delta(p,a,q)$ is either $1$ or $0$, so the possibility degrees of
existing transitions are the same, but if we work with a fuzzy automaton, they may be different.

To describe what happens when we start in any state and follow any sequence of events, we define
inductively an {\it extended fuzzy transition function} from $Q\times E^*\times Q$ to $[0,1]$,
denoted by the same notation $\delta$, as follows:

\begin{displaymath}
\delta(p,\epsilon,q)=\left\{ \begin{array}{ll}
1, & \textrm{if $q=p$}\\
0, & \textrm{otherwise}\qquad\qquad
\end{array} \right.
\end{displaymath}
$$\delta(p,\omega a,q)=\vee_{r\in
Q}(\delta(p,\omega,r)\wedge\delta(r,a,q))
$$
for all $\omega\in E^{*}$ and $a\in E$.

The reason we use the maxmin rule in the previous definition is that the state transition from $p$
to $q$ in a fuzzy process may be considered as water flow (gas, electricity, traffic flow, etc.)
through a water supply system of which the water pipes are series-parallel inter-connected.

Further, the language $\LL(G)$ generated by $G$, called {\it fuzzy language}, is defined as a fuzzy
subset of $E^*$ and given by
$$\LL(G)(\omega)=\vee_{q\in Q}\delta(q_0,\omega,q).$$
This means that a string $\omega$ of $E^{*}$ is not necessarily either ``in the fuzzy language
$\LL(G)$" or ``not in the fuzzy language $\LL(G)$"; rather $\omega$ has a membership grade
$\LL(G)(\omega)$, which measures its degree of membership in $\LL(G)$.

Without loss of generality, in this paper all automata are considered to be {\it accessible}, that
is, all states are reachable from the initial state.

If the state set of a fuzzy automaton $G$ is empty, then necessarily $\LL(G)=\OO$, that is,
$\LL(G)(\omega)=0$ for all $\omega\in E^*$; otherwise, it is obvious by the definition that
$\LL(G)$ has the following properties:

P1) $\LL(G)(\epsilon)=1;$

P2) $\LL(G)(\mu)\geq \LL(G)(\mu\nu)$ for any $\mu,\nu\in E^*.$

Conversely, given a fuzzy language $\LL$ (over the event set $E$) satisfying the above properties,
we can construct a fuzzy automaton $G=(Q_\LL,E_\LL,\delta_\LL,q_\LL)$, where

$Q_\LL=supp(\LL)$, $E_\LL=E$, $q_\LL=\epsilon$, and
\begin{displaymath}
\delta_\LL(\mu,a,\nu)=\left\{\begin{array}{ll}
\LL(\nu), & \textrm{if $\nu=\mu a$}\qquad\qquad\qquad\qquad\\
0, & \textrm{otherwise}
\end{array} \right.
\end{displaymath} for all $\mu,\nu\in supp(\LL)$ and $a\in E$.  There is no difficulty to verify
 that $\LL(G)=\LL$.

Note that in the rest of the paper, by a fuzzy language we mean the empty language $\OO$ or a fuzzy
language satisfying the above properties P1) and P2), unless otherwise specified. In particular,
the supports of such fuzzy languages are prefix closed by the property P2). We use $\F\LL$ to
denote the set of all fuzzy languages over $E$. More explicitly,
$\F\LL=\{\A\in\F(E^*):\A=\OO\textrm{ or }\A \textrm{ satisfies P1) and P2)}\}$.

Union and intersection of fuzzy languages can be defined as usual:
$$(\bigcup\limits_{i\in I}\LL_i)(\omega)=\vee_{i\in I}\LL_i(\omega), \mbox{ for all $\omega\in E^*$,
and}$$
$$(\bigcap\limits_{i\in I}\LL_i)(\omega)=\wedge_{i\in I}\LL_i(\omega), \mbox{ for all $\omega\in E^*$}.\qquad$$

The concatenation $\LL_1\LL_2$ of two fuzzy languages $\LL_1$ and $\LL_2$ is defined by
$$(\LL_1\LL_2)(\omega)=\vee\{\LL_1(\mu)\wedge\LL_2(\nu):\mu,\nu\in E^*\mbox{ and
}\mu\nu=\omega\},$$ for all $\omega\in E^*$. We sometimes write $(\LL_1\LL_2)(\omega)$ as
$\vee_{\mu\nu=\omega}(\LL_1(\mu)\wedge \LL_2(\nu))$ for convenience.

The following properties of fuzzy languages play an important role in the sequel.
\begin{Prop}If $\LL_1$ and $\LL_2$ are fuzzy languages, then $\LL_1\cup
\LL_2$, $\LL_1\cap \LL_2$, and $\LL_1\LL_2$ are fuzzy languages.\label{PFLan}
\end{Prop}
\begin{proof}If $\LL_1=\OO$ or $\LL_2=\OO$, then the proposition
holds clearly.

Assuming now that $\LL_1$ and $\LL_2$ are not empty, we show that $\LL_1\cup \LL_2$, $\LL_1\cap
\LL_2$, and $\LL_1\LL_2$ satisfy the properties P1) and P2), respectively. By the definitions, it
is obvious that $(\LL_1\cup\LL_2)(\epsilon)=(\LL_1\cap\LL_2)(\epsilon)= (\LL_1\LL_2)(\epsilon)=1$.

For any $\mu,\nu\in E^*$, we have that
\begin{eqnarray*}
 (\LL_1\cup\LL_2)(\mu\nu)
 &=&\LL_1(\mu\nu)\vee\LL_2(\mu\nu)
\\
 &\leq&\LL_1(\mu)\vee\LL_2(\mu)
 \\&=&(\LL_1\cup\LL_2)(\mu),
 \end{eqnarray*}and
\begin{eqnarray*}
 (\LL_1\cap\LL_2)(\mu\nu)
 &=&\LL_1(\mu\nu)\wedge\LL_2(\mu\nu)
\\
 &\leq&\LL_1(\mu)\wedge\LL_2(\mu)
 \\&=&(\LL_1\cap\LL_2)(\mu).
 \end{eqnarray*}

For $\LL_1\LL_2$, we have that $(\LL_1\LL_2)(\mu\nu)=\vee_{st=\mu\nu}(\LL_1(s)\wedge \LL_2(t))$ by
definition. Since $|\mu\nu|$ is finite, $\vee_{st=\mu\nu}(\LL_1(s)\wedge \LL_2(t))$ is the maximum
of $\{\LL_1(s)\wedge\LL_2(t):s,t\in E^*\mbox{ and }st=\mu\nu\}$, which is denoted by
$\LL_1(s_0)\wedge \LL_2(t_0)$. To prove $(\LL_1\LL_2)(\mu)\geq(\LL_1\LL_2)(\mu\nu)$, it is
sufficient to find out $s'$ and $t'$ that satisfy $s't'=\mu$ and $\LL_1(s')\wedge
\LL_2(t')\geq\LL_1(s_0)\wedge \LL_2(t_0)$, since $(\LL_1\LL_2)(\mu)=\vee_{s't'=\mu}(\LL_1(s')\wedge
\LL_2(t'))$. If $s_0\leq\mu$, say $s_0s_1=\mu$, then $s_1\leq t_0$. Taking $s'=s_0$ and $t'=s_1$,
we get that $s't'=\mu$ and $\LL_1(s')\wedge \LL_2(t')\geq\LL_1(s_0)\wedge \LL_2(t_0)$. If $s_0$ is
not a prefix of $\mu$, then this means that $\mu\leq s_0$. Then take $s'=\mu$ and $t'=\epsilon$. We
can also get that $s't'=\mu$ and $\LL_1(s')\wedge \LL_2(t')\geq\LL_1(s_0)\wedge \LL_2(t_0)$, thus
finishing the proof.
\end{proof}

For later need, we make the following observation that follows directly from definitions.
\begin{Prop}\

1) \ $supp(\LL_1\cup\LL_2)=\ supp(\LL_1)\cup supp(\LL_2);$

2) \ $supp(\LL_1\cap\LL_2)=\ supp(\LL_1)\cap supp(\LL_2);$

3) \ $supp(\LL_1\LL_2)=\ supp(\LL_1)supp(\LL_2).$\label{Psupp}
\end{Prop}
\begin{proof}The proof is evident and thus may be omitted.
\end{proof}

In crisp DES, one can build the overall model by building models of individual components first and
then composing them by product or parallel composition. We end this section by extending these
operations to fuzzy automata.

Given two fuzzy automata $G_i=(Q_i,E_i,\delta_i,q_{0i}),\ i=1,2$, we define two operations on them:
the product $G_1\times G_2$ and the parallel composition $G_1\parallel G_2$. These operations model
two forms of joint behavior of a set of fuzzy automata and we can think of them as two types of
systems resulting from the interconnection of system components $G_1$ and $G_2$.

The {\it product} of $G_1$ and $G_2$ is the fuzzy automaton

$G_1\times G_2=(Q_1\times Q_2,E_1\cap E_2,\delta,(q_{01},q_{02}))$,\\ where

$\delta((p_1,q_1),a,(p_2,q_2))=\delta_1(p_1,a,p_2)\wedge\delta_2(q_1,a,q_2)$ for all $(p_i,q_i)\in
Q_1\times Q_2 $ and $a\in E_1\cap E_2$.

As for crisp automata, it is easily verified that $\LL(G_1\times G_2)=\LL(G_1)\cap\LL(G_2).$ This
shows that the intersection of two fuzzy languages can be implemented by doing the product of their
automaton representations.

The {\it parallel composition} of $G_1$ and $G_2$ is the fuzzy automaton

 $G_1\parallel G_2=(Q_1\times Q_2,E_1\cup E_2,\delta,(q_{01},q_{02}))$,\\ where
$\delta((p_1,q_1),a,(p_2,q_2))=$
\begin{displaymath}
\left\{\begin{array}{ll}
\delta_1(p_1,a,p_2)\wedge\delta_2(q_1,a,q_2),&\textrm{if $\delta_1(p_1,a,p_2)>0$}\\\  & \textrm{\ \ and
$\delta_2(q_1,a,q_2)>0$}\\
\delta_1(p_1,a,p_2), & \textrm{if $a\in E_1\backslash E_2$ and $q_1=q_2$}\\
\delta_2(q_1,a,q_2), & \textrm{if $a\in E_2\backslash E_1$ and $p_1=p_2$}\\
0, & \textrm{otherwise.}
\end{array} \right.
\end{displaymath}
If $E_1=E_2$, then the parallel composition reduces to the product, since all fuzzy transitions are
forced to be synchronized.

\section{Controllability of Fuzzy Languages}
In this section, we first define fuzzy supervisor and the controllability of fuzzy languages. Based
on the two basic definitions, we then study what controlled behavior can be achieved when
controlling a fuzzy DES with a fuzzy supervisor. Some properties of controllable fuzzy languages
are also included in the section.

Our fuzzy DES model as described so far is simply a spontaneous generator of fuzzy event strings
without a means of external control. In order to control a fuzzy DES, we postulate that certain
events of the system can be disabled (i.e., prevented from occurring) with any degree when desired.
This enables us to influence the evolution of the system by preventing from the occurrence degrees
of some events at certain times. To model such control we partition the event set $E$ into {\it
controllable} and {\it uncontrollable} events: $E=E_c\dot\cup E_{uc}$, as usually done in crisp
DES. However, unlike controllable events in crisp DES, a controllable event in fuzzy DES can be
disabled with any degree. In other words, we allow that a controllable event in fuzzy DES can occur
incompletely. But uncontrollable events can never be disabled.

Control is achieved by means of a fuzzy supervisor, which is allowed to disable any fuzzy sets of
controllable events after having observed an arbitrary string $s\in supp(\LL(G))$. Formally, a {\it
fuzzy supervisor} for $G$ is a map $S:supp(\LL(G))\rightarrow\F(E)$ satisfying $S(s)(a)=1$ for any
$s\in supp(\LL(G))$ and $a\in E_{uc}$. The controlled system is denoted by $S/G$; the behavior of
$S/G$ is the fuzzy language $\LL^S$ obtained inductively as follows:

1) $\LL^S(\epsilon)=1;$

2) $\LL^S(sa)=\LL(G)(sa)\wedge S(s)(a)\wedge\LL^S(s)$ for any $s\in E^*$ and $a\in E$.

The empty string $\epsilon$ is always in $supp(\LL^S)$ since it is always contained in
$supp(\LL(G))$; here we have excluded the degenerate case where $G$ is the empty automaton.
Clearly, it follows from the definition of $\LL^S$ that $\LL^S\subseteq\LL(G)$ and $\LL^S\in\F\LL$.

Before giving the key definition, we have to introduce a fuzzy subset $\E_{uc}$ of $E^*$ which is
given by
\begin{displaymath}
\E_{uc}(\omega)=\left\{ \begin{array}{ll}
1, & \textrm{if $\omega\in E_{uc}$}\\
0, & \textrm{if $\omega\in E^*\backslash E_{uc}$.}
\end{array} \right.
\end{displaymath}

\begin{Def}A fuzzy language $\K\subseteq\LL(G)$ is said to be {\it controllable} with respect to
$\LL(G)$ and $E_{uc}$ if $$\K\E_{uc}\cap\LL(G)\subseteq\K.$$\label{Dfcont}
\end{Def}

The last expression reminds us of $\overline{K}E_{uc}\cap L(G)\subseteq\overline{K}$ in the
controllability definition of crisp languages. Indeed, if all the membership grades in $\K$ and
$\LL(G)$ are either $0$ or $1$, then the two expressions are the same, since the supports of $\K$
is prefix closed.

It is clear that the fuzzy languages $\OO$ and $\LL(G)$ are always controllable with respect to
$\LL(G)$ and any $E_{uc}$. Unless otherwise specified, controllability will always be with respect
to $\LL(G)$ and $E_{uc}$, and we simply call $\K$ controllable if the context is clear. Sometimes,
we also write $\LL$ for $\LL(G)$. For convenience, we will exclude the special discussions on the
empty fuzzy language in the proofs of our results since $\OO$ is trivial and the results are
obvious.

We first make the following observation from Definition \ref{Dfcont}.
\begin{Prop}A fuzzy language $\K$ is controllable if and only if $\K(s)\wedge\LL(G)(sa)=\K(sa)$ for any
$s\in E^*$ and $a\in E_{uc}$.\label{Puc}
\end{Prop}
\begin{proof}Suppose that $\K$ is controllable. Using the controllability definition and $a\in E_{uc}$, we have that
$\K(s)\wedge\LL(G)(sa)=(\K\E_{uc}\cap\LL(G))(sa)\leq\K(sa)$. On the other hand, it is clear that
$\K(sa)\leq\K(s)\wedge\LL(G)(sa)$. This yields that $\K(s)\wedge\LL(G)(sa)=\K(sa)$.

Conversely, assuming that $\K(s)\wedge\LL(G)(sa)=\K(sa)$ for any $s\in E^*$ and $a\in E_{uc}$, we
need prove that $\K\E_{uc}\cap\LL(G)\subseteq\K$, namely
$(\K\E_{uc}\cap\LL(G))(\omega)\leq\K(\omega)$ for any $\omega\in E^*$. If $|\omega|=0$, namely
$\omega=\epsilon$, then it is obvious that $(\K\E_{uc}\cap\LL(G))(\omega)=0\leq\K(\omega)$. In the
case of $|\omega|>0$, we can write $\omega$ as $\tilde{\omega}a$ for some $a\in E$. Further, if
$a\in E_{c}$, then it is also obvious that $(\K\E_{uc}\cap\LL(G))(\omega)=0\leq\K(\omega)$;
otherwise, by assumption we have that
$(\K\E_{uc}\cap\LL(G))(\omega)=\K(\tilde{\omega})\wedge\LL(G)(\tilde{\omega}a)
=\K(\tilde{\omega}a)=\K(\omega)$. Overall, we get that
$(\K\E_{uc}\cap\LL(G))(\omega)\leq\K(\omega)$ for any $\omega\in E^*$, thereby finishing the proof.
\end{proof}

Proposition \ref{Puc} thus provides an equivalent definition of the controllability of fuzzy
languages.

The following theorem shows us when a desired fuzzy language can be synthesized through a fuzzy
supervisor.

\begin{Thm}Let $\K\subseteq\LL(G)$, where $\K$ is a nonempty fuzzy language. Then there exists a
fuzzy supervisor $S$ for $G$ such that $\LL^S=\K$ if and only if $\K$ is controllable.\label{Tcont}
\end{Thm}
\begin{proof}First we prove the sufficiency by constructing a desired fuzzy supervisor. For any $s\in
supp(\LL(G))$, define $S(s)$ according to
\begin{displaymath}
S(s)(a)=\left\{ \begin{array}{ll}
1, & \textrm{if $a\in E_{uc}$}\\
\K(sa), & \textrm{if $a\in E_{c}$}
\end{array} \right.
\end{displaymath} for all $a\in E$. Obviously, $S$ is a fuzzy supervisor.

We claim that $\LL^S=\K$, which will be proved by induction on the length of the strings $s$.

$\bullet$ The basis step is for strings of length $0$, namely, $s=\epsilon$. By assumption, $\K$ is
a nonempty fuzzy
    language, so $\K(\epsilon)=1$ by definition. On the other hand,
    we always have that $\LL^S(\epsilon)=1$ by definition. Thus the basis step is true.

$\bullet$ Assume that for all strings $s$ with $|s|\leq n$, we have that
    $\LL^S(s)=\K(s)$. We now prove the same for strings of the form $sa$,
    where $a\in E$. If $s\not\in supp(\LL(G))$, then $\LL(G)(s)=0$. Therefore
     $\LL(G)(sa)=0$ by the definition of fuzzy languages. Noting that
     $\LL^S\subseteq\LL(G)$ and $\K\subseteq\LL(G)$, we obtain
     that $\LL^S(sa)=\K(sa)=0$. If $s\in supp(\LL(G))$, then
     we can define $S(s)$ as above, and two cases arise.

    Case 1: $a\in E_{uc}$. In this case, we get that $S(s)(a)=1$ by definition. Thus,
    \begin{eqnarray*}
 \LL^S(sa)&=&
    \LL(G)(sa)\wedge S(s)(a)\wedge\LL^S(s)\ \mbox{(by definition)}\\&=&
    \LL(G)(sa)\wedge\LL^S(s)\\&=&\LL(G)(sa)\wedge\K(s)
    \ \ \mbox{(by induction hypothesis)}\\
    &=&\K(sa)
    \ \ \mbox{(by Proposition \ref{Puc}),}
 \end{eqnarray*}i.e., $\LL^S(sa)=\K(sa)$.

Case 2: $a\in E_{c}$. Using the definition of $S(s)$ and the induction hypothesis, we have the
following:\begin{eqnarray*} \LL^S(sa)
 &=&\LL(G)(sa)\wedge S(s)(a)\wedge\LL^S(s) \\
 &=&\LL(G)(sa)\wedge \K(sa)\wedge\LL^S(s)
 \\&=&\LL(G)(sa)\wedge \K(sa)\wedge\K(s)
 \\&=&\LL(G)(sa)\wedge \K(sa)
 \\&=&\K(sa),
\end{eqnarray*}that is, $\LL^S(sa)=\K(sa)$. The claim has been proved, thus finishing
the proof of the sufficiency.

Next, to see the necessity, suppose that there is a fuzzy supervisor $S$ for $G$ such that
$\LL^S=\K$. To show the controllability of $\K$, by Proposition \ref{Puc} it suffices to prove that
$\LL^S(s)\wedge\LL(G)(sa)=\LL^S(sa)$ for any $s\in E^*$ and $a\in E_{uc}$. In fact, by definition
we have that $\LL^S(sa)=\LL(G)(sa)\wedge S(s)(a)\wedge\LL^S(s)=\LL^S(s)\wedge\LL(G)(sa)$ since
$S(s)(a)=1$ for $a\in E_{uc}$. This completes the
 proof of the theorem.
\end{proof}

\begin{Rem}From the proof of the sufficiency of Theorem \ref{Tcont}, we find that such a fuzzy supervisor is not unique
in general. That is, if there exists a fuzzy supervisor, say $S_0$, such that $\LL^{S_0}=\K$, then
there are generally more than one fuzzy supervisors satisfying the requirements. In fact, more
fuzzy supervisors can be constructed by
\begin{displaymath}
S(s)(a)=\left\{ \begin{array}{ll}
1, & \textrm{if $a\in E_{uc}$}\\
\K(sa), & \textrm{if $a\in E_{c}$ and}\\\  & \textrm{
$\LL(sa)\wedge\K(s)>\K(sa)$}\\
x\in[\K(sa),1], & \textrm{if $a\in E_{c}$ and}\\\  & \textrm{ $\LL(sa)\wedge\K(s)=\K(sa).$}
\end{array} \right.
\end{displaymath}
Imitating the above proof of the sufficiency, we can easily prove that $\LL^S=\K$.
\end{Rem}

For the sake of illustrating the above theorem and remark, let us consider a very simple example.
\begin{Eg}Let $E=\{a,b\}$ and $E_{uc}=\{b\}$. $\LL$ and $\K$ are given by
$$\LL=\frac{1}{\epsilon}+\frac{0.9}{a}+\frac{0.7}{aa}+\frac{0.7}{ab}+\frac{0.5}{aba},$$
$$\K=\frac{1}{\epsilon}+\frac{0.8}{a}+\frac{0.7}{aa}+\frac{0.7}{ab}.\quad\qquad$$ Take a fuzzy supervisor $S$ as
follows: $$S(\epsilon)=\frac{0.8}{a}+\frac{1}{b},\ \ S(a)=\frac{0.7}{a}+\frac{1}{b},$$
$$S(aa)=S(ab)=S(aba)=\frac{0}{a}+\frac{1}{b}.$$
Then we see that $\LL^S=\K$. In fact, we can also achieve this aim by taking the following fuzzy
supervisor:
$$S(\epsilon)=\frac{0.8}{a}+\frac{1}{b},\ S(a)=\frac{x}{a}+\frac{1}{b},\ S(aa)=\frac{y}{a}+\frac{1}{b},$$
$$S(ab)=\frac{0}{a}+\frac{1}{b},\ \ \ S(aba)=\frac{z}{a}+\frac{1}{b},$$where
$x\in[0.7,1]$ and $y,z\in[0,1].$
\end{Eg}

Based on Theorem \ref{Tcont}, we can now formulate the following Supervisory Control Problem (SCP)
for fuzzy DES:

SCP: Given fuzzy DES $G$ with event set $E$, uncontrollable event set $E_{uc}\subseteq E$, and two
fuzzy languages $\LL_a$ and $\LL_l$, where $\OO\neq\LL_a\subseteq\LL_l\subseteq\LL(G)$, find a
fuzzy supervisor $S$ such that $\LL_a\subseteq\LL^{S}\subseteq\LL_l$.

Here, $\LL_a$ describes the minimal acceptable behavior and $\LL_l$ describes the maximal legal
behavior. SCP requires to find a fuzzy supervisor such that the behavior of controlled system is
both acceptable and legal. A complete solution to this problem will be given at the end of Section
VI.

The following fact shows that controllable fuzzy languages are closed under arbitrary unions and
intersections, respectively.
\begin{Prop}If fuzzy languages $\K_i$, $i\in I $, are controllable, then so are
$\bigcup\limits_{i\in I}\K_i$ and $\bigcap\limits_{i\in I}\K_i$.\label{PContr}
\end{Prop}
\begin{proof}For simplicity, we prove the case that $I$ is finite. There is no difficulty to generalize
 the proof to infinite index set. If $\K_1$ and $\K_2$ are controllable fuzzy languages, then by Proposition \ref{PFLan}
  we know that $\K_1\cup\K_2$ and $\K_1\cap\K_2$ are fuzzy languages. We first show that $\K_1\cup\K_2$ is controllable.
For any $s\in E^*$ and $a\in E_{uc}$, we have that
\begin{eqnarray*}
&\ &(\K_1\cup\K_2)(s)\wedge\LL(sa) \\
 &=&(\K_1(s)\vee\K_2(s))\wedge\LL(sa)
 \\&=&(\K_1(s)\wedge\LL(sa))\vee(\K_2(s)\wedge\LL(sa))
 \\&=&\K_1(sa)\vee\K_2(sa)\ \mbox{ (by controllability)}
 \\&=&(\K_1\cup\K_2)(sa),
 \end{eqnarray*}that is, $(\K_1\cup\K_2)(s)\wedge\LL(sa)=(\K_1\cup\K_2)(sa)$. Hence,
 $\K_1\cup\K_2$ is controllable by Proposition \ref{Puc}.
 Similarly, one can
 prove that $\K_1\cap\K_2$ is controllable. Thus the proposition follows when $I$ is finite.
\end{proof}

Let us turn now to discussing what can be done when the given fuzzy language $\K$ is
uncontrollable. If $\K$ is not controllable, then it is natural to consider the possibility of
approximating $\K$ by some controllable fuzzy languages. As in crisp DES, two questions arise: Is
there the ``largest" controllable fuzzy sublanguage of $\K$, and is there the ``least" controllable
fuzzy superlanguage of $\K$, where ``largest" and ``least" are in terms of fuzzy set inclusion.
Both answers are positive, and we address the two questions in the subsequent two sections,
respectively.

For later need, we define the class of controllable fuzzy sublanguages of $\K$ and the class of
controllable fuzzy superlanguages of $\K$ as follows:
$$\mathscr{C}_{sub}(\K)=\{\M\in\F\LL:\M\subseteq\K\mbox{ and $\M$ is controllable}\};$$
$$\mathscr{C}_{sup}(\K)=\{\M\in\F\LL:\K\subseteq\M\subseteq\LL\mbox{ and $\M$ is controllable}\}.$$
Observe that $\OO\in\mathscr{C}_{sub}(\K)$ and $\LL\in\mathscr{C}_{sup}(\K)$, so the two classes
are not empty.

\section{Supremal Controllable Fuzzy Sublanguage}
This section is devoted to finding the ``largest" controllable fuzzy sublanguage of $\K$ mentioned
above and discussing its calculation when $\K$ is not controllable.

Define $\K^\uparrow=\bigcup\limits_{\M\in\mathscr{C}_{sub}(\K)}\M$. Then the existence of the
largest controllable fuzzy sublanguage of $\K$ follows from the following.

\begin{Prop}The fuzzy language $\K^\uparrow$ is
the largest controllable fuzzy sublanguage of $\K$.\label{PLargCon}
\end{Prop}
\begin{proof}
The assertion follows directly from Proposition \ref{PContr}.
\end{proof}

Since by definition $\M\subseteq\K^\uparrow$ for any $\M\in\mathscr{C}_{sub}(\K)$, we call
$\K^\uparrow$ the {\it supremal controllable fuzzy sublanguage} of $\K$. If $\K$ is controllable,
then $\K^\uparrow=\K$. In the ``worst" case, $\K^\uparrow=\OO$. We will refer to ``$\uparrow$" as
the operation of obtaining the supremal controllable fuzzy sublanguage.

Several useful properties of the operation are presented in the following proposition.

\begin{Prop}Let $\K_1,\K_2\in\F\LL$.

1) \ If $\K_1\subseteq\K_2$, then $\K_1^\uparrow\subseteq\K_2^\uparrow$.

2) \ $(\K_1\cap\K_2)^\uparrow=\K_1^\uparrow\cap\K_2^\uparrow$.

3) \ $\K_1^\uparrow\cup\K_2^\uparrow\subseteq(\K_1\cup\K_2)^\uparrow$.
\end{Prop}
\begin{proof}

1) \ It is clear by the definition of the operation $\uparrow$.

2) \ Using 1), we get that $(\K_1\cap\K_2)^\uparrow\subseteq\K_1^\uparrow$ and
$(\K_1\cap\K_2)^\uparrow\subseteq\K_2^\uparrow$, so
$(\K_1\cap\K_2)^\uparrow\subseteq\K_1^\uparrow\cap\K_2^\uparrow$. By contradiction, suppose that
$(\K_1\cap\K_2)^\uparrow\varsubsetneq\K_1^\uparrow\cap\K_2^\uparrow$. Then there exists $s\in E^*$
such that $(\K_1\cap\K_2)^\uparrow(s)<\K_1^\uparrow(s)\wedge\K_2^\uparrow(s)$, that is,
$\K_i^\uparrow(s)>(\K_1\cap\K_2)^\uparrow(s)$ for $i=1,2$. By definition,
$\K_i^\uparrow(s)=\vee\{\M_{i}(s):\M_{i}\in\mathscr{C}_{sub}(\K_i)\}$ and
$(\K_1\cap\K_2)^\uparrow(s)=\vee\{\M(s):\M\in\mathscr{C}_{sub}(\K_1\cap\K_2)\}$. Therefore there is
at least one $\M'_{i}\in\mathscr{C}_{sub}(\K_i)$ such that $\M'_i(s)>\M(s)$ for all
$\M\in\mathscr{C}_{sub}(\K_1\cap\K_2)$. Otherwise, it contradicts with
$\K_i^\uparrow(s)>(\K_1\cap\K_2)^\uparrow(s)$. Hence $(\M'_1\cap\M'_2)(s)>\M(s)$ for all
$\M\in\mathscr{C}_{sub}(\K_1\cap\K_2)$. Observe that
$\M'_1\cap\M'_2\in\mathscr{C}_{sub}(\K_1\cap\K_2)$ since $\M'_1\cap\M'_2\subseteq\K_1\cap\K_2$ and
$\M'_1\cap\M'_2$ is controllable by Proposition \ref{PContr}. Taking $\M=\M'_1\cap\M'_2$, then we
get $(\M'_1\cap\M'_2)(s)>(\M'_1\cap\M'_2)(s)$, which is absurd. Thus
$(\K_1\cap\K_2)^\uparrow=\K_1^\uparrow\cap\K_2^\uparrow$.

3) \ By 1), we obtain that $\K_1^\uparrow\subseteq(\K_1\cup\K_2)^\uparrow$ and
$\K_2^\uparrow\subseteq(\K_1\cup\K_2)^\uparrow$, so
$\K_1^\uparrow\cup\K_2^\uparrow\subseteq(\K_1\cup\K_2)^\uparrow$.
\end{proof}

\begin{Rem}In general, $\K_1^\uparrow\cup\K_2^\uparrow=(\K_1\cup\K_2)^\uparrow$ cannot hold. For
example, let $$E=E_{uc}=\{a,b\},\ \LL=\frac{1}{\epsilon}+\frac{0.8}{a}+\frac{0.5}{b},$$
$$\K_1=\frac{1}{\epsilon}+\frac{0.7}{a}+\frac{0.5}{b}, \mbox{ and }
\K_2=\frac{1}{\epsilon}+\frac{0.8}{a}+\frac{0.4}{b}.$$ Then
$\K_1^\uparrow\cup\K_2^\uparrow=\OO\cup\OO=\OO$, whereas
$(\K_1\cup\K_2)^\uparrow=\LL^\uparrow=\LL$.
\end{Rem}

Next, we wish to give an explicit expression of the membership function associated with
$\K^\uparrow$. For this purpose, let us first define inductively an auxiliary fuzzy subset
$\hat{\K}$ of $E^*$:

\begin{displaymath}
\hat{\K}(\epsilon)=\left\{ \begin{array}{ll}
0, & \textrm{if $\K(\omega)<\LL(\omega)$ for some $\omega\in E^*_{uc}$}\\
1, & \textrm{otherwise};
\end{array} \right.
\end{displaymath}
\begin{displaymath}
\hat{\K}(sa)=\left\{ \begin{array}{ll}
\hat{\K}(s)\wedge\K(sa), & \textrm{if $a\in E_{uc}$\quad\quad}\\
\hat{\K}(s)\wedge\K(sa)\wedge\lambda(sa), & \textrm{if $a\in E_{c}$}
\end{array} \right.
\end{displaymath}for all $s\in E^*$ and $a\in E$, where we
use $\lambda(sa)$ to denote $\wedge\{\K(t):t\in saE_{uc}^*\mbox{ and
}\K(t)<\LL(t)\wedge\K(\tilde{t})\}$ and use the abbreviation $saE_{uc}^*$ for $\{s\}\{a\}E_{uc}^*$.
Recall that for any $s\in E^*$ with $|s|\geq1$, we use $\tilde{s}$ to denote the maximal proper
prefix of $s$. If $\{\K(t):t\in saE_{uc}^*\mbox{ and }\K(t)<\LL(t)\wedge\K(\tilde{t})\}=\emptyset$,
then we set $\lambda(sa)=1$.

Our aim now is to show that $\K^\uparrow=\hat{\K}$. To this end, it is convenient to have the
following lemmas.
\begin{Lem}$\hat{\K}\in\F\LL$.\label{LKisFLan}
\end{Lem}
\begin{proof}By the above definition, it is clear that $\hat{\K}\in\F(E^*)$. If
$\hat{\K}(\epsilon)=0$, then by induction on the length of strings we can easily get that
$\hat{\K}(s)=0$ for any $s\in E^*$. Thus $\hat{\K}$ is the empty fuzzy language when
$\hat{\K}(\epsilon)=0$. In the case of $\hat{\K}(\epsilon)=1$, note that by definition we always
have $\hat{\K}(sa)\leq\hat{\K}(s)$. This implies that $\hat{\K}$ satisfies the property P2). Hence
we also have that $\hat{\K}\in\F\LL$ when $\hat{\K}(\epsilon)=1$, finishing the proof.
\end{proof}

\begin{Lem}$\hat{\K}(s)\wedge\K(sa)=\hat{\K}(s)\wedge\LL(sa)$ for any $s\in E^*$ and $a\in
E_{uc}$.\label{LK=L}
\end{Lem}
\begin{proof}If $\hat{\K}(\epsilon)=0$, then we have that $\hat{\K}=\OO$ from the proof of Lemma
\ref{LKisFLan}. So $\hat{\K}(s)\wedge\K(sa)=\hat{\K}(s)\wedge\LL(sa)=0$.

Let us consider the other case $\hat{\K}(\epsilon)=1$. Two subcases need to be discussed:

Subcase 1: $s\in E^*_{uc}$. Since $\hat{\K}(\epsilon)=1$, we know by definition that
$\K(\omega)=\LL(\omega)$ for all $\omega\in E^*_{uc}$. Thus $\K(sa)=\LL(sa)$ because of $a\in
E_{uc}$. Consequently, $\hat{\K}(s)\wedge\K(sa)=\hat{\K}(s)\wedge\LL(sa)$.

Subcase 2: $s\not\in E^*_{uc}$. In this subcase, $s$ can be written uniquely as $s=\mu c\nu$ for
some $\mu\in E^*$, $c\in E_c$, and $\nu\in E^*_{uc}$.

If $|\nu|=0$, namely, $s=\mu c$, then by definition we have the following:
\begin{eqnarray*}
\hat{\K}(s)\wedge\K(sa)
 &=&\hat{\K}(\mu c)\wedge\K(\mu ca)\\
 &=&\hat{\K}(\mu)\wedge\K(\mu c)\wedge\lambda(\mu c)\wedge\K(\mu ca)
=:(1)
\end{eqnarray*}
\begin{eqnarray*}
\hat{\K}(s)\wedge\LL(sa) &=&\hat{\K}(\mu c)\wedge\LL(\mu ca)\\&=&\hat{\K}(\mu)\wedge\K(\mu
c)\wedge\lambda(\mu c)\wedge\LL(\mu ca)=:(2).
\end{eqnarray*}
Clearly, $\K(\mu ca)\leq\LL(\mu ca)\wedge\K(\mu c)$. When $\K(\mu ca)=\LL(\mu ca)\wedge\K(\mu c)$,
we get that $$(1)=\hat{\K}(\mu)\wedge\K(\mu c)\wedge\lambda(\mu c)\wedge\LL(\mu ca)=(2).$$If
$\K(\mu ca)<\LL(\mu ca)\wedge\K(\mu c)$, then
\begin{eqnarray*}
\lambda(\mu c)
 &=&\wedge\{\K(t):t\in\mu cE_{uc}^*\mbox{ and }\K(t)<\LL(t)\wedge\K(\tilde{t})\}\\
 &\leq&\K(\mu ca)\\
&\leq&\LL(\mu ca).
\end{eqnarray*}
Hence,
$$(1)=\hat{\K}(\mu)\wedge\K(\mu c)\wedge\lambda(\mu c)=(2).$$

If $|\nu|>0$, say $|\nu|=n$, then we have that $\nu=\nu_1\nu_2\cdots\nu_n$ for some $\nu_i\in
E_{uc}$. Thus by definition we obtain that$$\hat{\K}(s)=\hat{\K}(\mu
c\nu_1\nu_2\cdots\nu_n)=\hat{\K}(\mu c)\wedge\K(s).$$ So we get that
\begin{eqnarray*}
\hat{\K}(s)\wedge\K(sa)
 &=&\hat{\K}(\mu c)\wedge\K(s)\wedge\K(sa)\\
 &=&\hat{\K}(\mu)\wedge\K(\mu c)\wedge\lambda(\mu c)\wedge\K(s)\wedge\K(sa)\\
 &=&\hat{\K}(\mu)\wedge\lambda(\mu c)\wedge\K(s)\wedge\K(sa)
=:(3)
\end{eqnarray*}and
\begin{eqnarray*}
\hat{\K}(s)\wedge\LL(sa)
 &=&\hat{\K}(\mu c)\wedge\K(s)\wedge\LL(sa)\\
 &=&\hat{\K}(\mu)\wedge\K(\mu c)\wedge\lambda(\mu c)\wedge\K(s)\wedge\LL(sa)\\
 &=&\hat{\K}(\mu)\wedge\lambda(\mu c)\wedge\K(s)\wedge\LL(sa)=:(4).
\end{eqnarray*}
If $\K(sa)=\LL(sa)\wedge\K(s)$, then we see that $$(3)=\hat{\K}(\mu)\wedge\lambda(\mu
c)\wedge\K(s)\wedge\LL(sa)=(4).$$If $\K(sa)<\LL(sa)\wedge\K(s)$, then
\begin{eqnarray*}
\lambda(\mu c)
 &=&\wedge\{\K(t):t\in\mu cE_{uc}^*\mbox{ and }\K(t)<\LL(t)\wedge\K(\tilde{t})\}\\
 &\leq&\K(\mu
c\nu_1\nu_2\cdots\nu_n a)\\
&=&\K(sa)\\
&\leq&\LL(sa).
\end{eqnarray*}
This gives rise to
$$(3)=\hat{\K}(\mu)\wedge\lambda(\mu c)\wedge\K(s)=(4).$$
Overall, we have that $\hat{\K}(s)\wedge\K(sa)=\hat{\K}(s)\wedge\LL(sa)$ for any $s\in E^*$ and
$a\in E_{uc}$, and the proof is completed.
\end{proof}

\begin{Lem}Suppose that $\M\subseteq\LL$ is controllable and $\M\neq\OO$. Then $\M(\mu)=\LL(\mu)$ for
any $\mu\in E^*_{uc}$.\label{LM=L}
\end{Lem}
\begin{proof}It is clear that $\M(\epsilon)=\LL(\epsilon)=1$ since $\M\neq\OO$. Assume that
$\M(\mu)=\LL(\mu)$ for all strings $\mu\in E^*_{uc}$ with $|\mu|\leq n$. For any $a\in E_{uc}$, we
have that
\begin{eqnarray*}
\M(\mu a)&=&\M(\mu)\wedge\LL(\mu a)\ \mbox{(by the controllability of $\M$)}\\
&=&\LL(\mu)\wedge\LL(\mu a)\ \mbox{(by induction hypothesis)}
\\&=&\LL(\mu a),
\end{eqnarray*}i.e., $\M(\mu a)=\LL(\mu a)$, finishing the proof.
\end{proof}

Based upon the above three lemmas, we now verify the fact below.
\begin{Thm}
$\K^\uparrow=\hat{\K}$.\label{Tuparrow}
\end{Thm}
\begin{proof}
Let us first show that $\hat{\K}\subseteq\K^\uparrow$. It suffices to verify that
$\hat{\K}\in\mathscr{C}_{sub}(\K)=\{\M\in\F\LL:\M\subseteq\K\mbox{ and $\M$ is controllable }\}$.
By Lemma \ref{LKisFLan}, we see that $\hat{\K}$ is a fuzzy language; moreover,
$\hat{\K}\subseteq\K$ by the definition of $\hat{\K}$. It remains only to show that $\hat{\K}$ is
controllable. For any $s\in E^*$ and $a\in E_{uc}$, we see that
$\hat{\K}(s)\wedge\LL(sa)=\hat{\K}(s)\wedge\K(sa)=\hat{\K}(sa)$ from Lemma \ref{LK=L}
 and the definition of $\hat{\K}$. Therefore, it follows from Proposition \ref{Puc} that
 $\hat{\K}$ is controllable.

For the reverse inclusion, take any $\M\in\mathscr{C}_{sub}(\K)$. It is enough to prove that
$\M\subseteq\hat{\K}$; we prove it by showing that $\M(s)\leq\hat{\K}(s)$ for any $s\in E^*$. Use
induction on the length of $s$.

In the basis step, namely, $s=\epsilon$, if $\M(\epsilon)=0$ or $\hat{\K}(\epsilon)=1$, then it is
obvious that $\M(\epsilon)\leq\hat{\K}(\epsilon)$. Suppose now that $\M(\epsilon)=1$ and
$\hat{\K}(\epsilon)=0$. By $\hat{\K}(\epsilon)=0$, we know that there exists $\mu\in E^*_{uc}$ such
that $\K(\mu)<\LL(\mu)$. So $\M(\mu)\leq\K(\mu)<\LL(\mu)$, which contradicts with Lemma \ref{LM=L}.
Hence we have that $\M(s)\leq\hat{\K}(s)$ in the basis step.

The induction hypothesis is that $\M(s)\leq\hat{\K}(s)$ for all strings $s$ satisfying $|s|\leq n$.
We now prove the same for strings of the form $sa$. By contradiction, let us suppose that
$\M(sa)>\hat{\K}(sa)$ for some $a\in E$. If $a\in E_{uc}$, then by definition and the induction
hypothesis we get that $\hat{\K}(sa)=\hat{\K}(s)\wedge\K(sa)\geq\M(s)\wedge\M(sa)=\M(sa)$. This
contradicts with the assumption that $\M(sa)>\hat{\K}(sa)$. If $a\in E_{c}$, then
$\hat{\K}(sa)=\hat{\K}(s)\wedge\K(sa)\wedge\lambda(sa)$ by definition. Since
$\hat{\K}(s)\geq\M(s)\geq\M(sa)$, $\K(sa)\geq\M(sa)$, and $\M(sa)>\hat{\K}(sa)$, we see that
$\hat{\K}(sa)=\lambda(sa)$. That is,
$$\hat{\K}(sa)=\wedge\{\K(t):t\in saE_{uc}^*\mbox{ and
}\K(t)<\LL(t)\wedge\K(\tilde{t})\}.$$Since $\M(sa)>\hat{\K}(sa)$, there by the definition of
infimum exists $t_0\in sa E^*_{uc}$ satisfying $\K(t_0)<\LL(t_0)\wedge\K(\tilde{t_0})$ such that
$\K(t_0)<\M(sa)$. Obviously, $t_0\neq sa$; otherwise, it contradicts with $\M\subseteq\K$. Thus we
can write $t_0=sa\nu_1\nu_2\cdots\nu_m$ for some $\nu_i\in E_{uc}$, where $m\geq1$. By the
controllability of $\M$, we see that
$\M(t_0)=\M(\tilde{t_0})\wedge\LL(t_0)=\M(\tilde{t_0})\wedge\K(\tilde{t_0})\wedge\LL(t_0)$. Using
repeatedly Proposition \ref{Puc}, we can obtain that
$\M(\tilde{t_0})=\M(sa)\wedge\LL(\tilde{t_0})$. Thus
$\M(t_0)=\M(sa)\wedge\LL(\tilde{t_0})\wedge\K(\tilde{t_0})\wedge\LL(t_0)
=\M(sa)\wedge\K(\tilde{t_0})\wedge\LL(t_0)$. Noting that $\M(sa)>\K(t_0)$ and
$\K(\tilde{t_0})\wedge\LL(t_0)>\K(t_0)$ by the previous arguments, we see that $\M(t_0)>\K(t_0)$,
which contradicts with $\M\subseteq\K$ again. The proof of the induction step is finished, and the
proof of the theorem is also completed.
\end{proof}

According to Theorem \ref{Tuparrow}, we obtain an explicit expression of the membership function
associated with $\K^\uparrow$, as promised; however, it seems to be difficult to give a compact
formula for $\K^\uparrow$ when $\K$ is not controllable. Here we provide an upper bound for
$supp(\K^\uparrow)$. Before doing this, we have to define the quotient of (crisp) languages. Let
$L_1,L_2\subseteq E^*$. The quotient operation $L_1/L_2$ is defined as follows:$$L_1/L_2=\{s\in
E^*:st\in L_1\mbox{ for some }t\in L_2\}.$$

\begin{Prop}\

$supp(\K^\uparrow)\subseteq supp(\K)\backslash[(supp(\LL)\backslash
supp(\K))/E^*_{uc}]E^*.$\label{PsuppK}
\end{Prop}
\begin{proof}If $supp(\K^\uparrow)=\emptyset$, namely, $\K^\uparrow=\OO$, then the proposition evidently
holds; otherwise, take any $s\in supp(\K^\uparrow)$. By contradiction, assume that $s\not\in
supp(\K)\backslash[(supp(\LL)\backslash supp(\K))/E^*_{uc}]E^*.$ This means that
$s\in[(supp(\LL)\backslash supp(\K))/E^*_{uc}]E^*$ since $s\in supp(\K^\uparrow)\subseteq
supp(\K)$. Thus there exist $s_1,s_2\in E^*,$ and $\mu\in E^*_{uc}$ such that $s=s_1s_2$ and
$s_1\mu\in supp(\LL)$, but $s_1\mu\not\in supp(\K)$. Obviously, $\mu\neq\epsilon$. Otherwise it
contradicts with the fact that $supp(\K)$ is prefix closed. So we may suppose that
$\mu=\mu_1\mu_2\cdots\mu_n$ for some $\mu_i\in E_{uc}$, where $n\geq1$. Further, using recursively
Proposition \ref{Puc}, we have that $\K^\uparrow(s_1\mu)=\K^\uparrow(s_1)\wedge\LL(s_1\mu).$ Noting
that $s_1\leq s\in supp(\K^\uparrow)$ and $s_1\mu\in supp(\LL)$, we see that
$\K^\uparrow(s_1\mu)>0$, which contradicts with $s_1\mu\not\in supp(\K)$. As $s$ was arbitrary, the
proposition has been proved.
\end{proof}

\begin{Rem}The upper bound for $supp(\K^\uparrow)$ is optimal in the sense that it can be achieved
by some fuzzy languages. However, the reverse inclusion in Proposition \ref{PsuppK} is not true in
general, as shown by the following example.
\end{Rem}
\begin{Eg}
Let $E=\{a,b,c\}$ and $E_{uc}=\{b\}$. $\LL$ and $\K$ are given by $$\LL(\epsilon)=\LL(a)=1,\
\LL(c)=0.5,\ \LL(ab^n)=1/n\mbox{ for any }n\geq1;$$
$$\K(\epsilon)=\K(a)=1,\ \K(ab^n)=1/(n+1)\mbox{ for any }n\geq1.$$ One can use the definition of
$\hat{\K}$ to get that $\K^\uparrow={1}/{\epsilon}$, so $supp(\K^\uparrow)=\{\epsilon\}$. However,
the right side of Proposition \ref{PsuppK} equals $supp(\K)=\{\epsilon,a,ab^n\}$, where $n\geq1$.
\end{Eg}

We next provide an algorithm to compute $\K^\uparrow$. For simplicity, we assume that $\K$ and
$\LL$ are deterministic regular for the moment, that is, $\K$ and $\LL$ are generated by
deterministic finite state fuzzy automata, respectively.

Let $G=(Q,E,\delta,q_0)$ and $H=(P,E,\delta_H,p_0)$ be deterministic finite state fuzzy automata
that generate $\LL$ and $\K$, respectively, i.e., $\LL(G)=\LL$ and $\LL(H)=\K$, where it is assumed
that $\K\subseteq\LL(G)$.

{\it Algorithm for $\K^\uparrow$:}

Step 1): Let $H_0=(P_0,E,\delta_{H_0},(p_0,q_0))=H\times G$. Thus $\LL(H_0)=\K$. States of $H_0$
will be denoted by pairs $(p,q)$. Let $i=0$.

Step 2): Compare $H_i$ with $G$. For each $(p,q)\in P_i$, if $\delta_{H_i}((p,q),a,(p',q'))=0$ for
some $(p',q')\in P\times Q$ and $a\in E_{uc}$, but $\delta(q,a,q')>0$, then remove $(p,q)$ and all
its associated transitions from $H_i$. Take the accessible operation to give rise to $H_{i+1}$.

Step 3): If no states were removed from $H_{i+1}$ in Step 2), go to Step 4). Else, set $i\leftarrow
i+1$ and go to Step 2).

Step 4): If $H_{i+1}$ is the empty automaton, then we set $H'=H_{i+1}$ and STOP.

If there is a path $\omega$ consisting of uncontrollable events and beginning with the initial
state $(p_0,q_0)$ of $H_{i+1}$ such that
$\delta_{H_{i+1}}((p_0,q_0),\omega,(p',q'))<\delta(q_0,\omega,q')$, then we let $H'$ be the empty
automaton and STOP, too.

Otherwise, we reassign membership grades of controllable events appearing in the transitions of
$H_{i+1}$. If $\delta_{H_{i+1}}((p_0,q_0),sa,(p',q'))>0$, where $s\in E^*$ and $a$ is a
controllable event and, say, connects $(p_s,q_s)$ and $(p',q')$, then set
$\delta'_{H_{i+1}}((p_s,q_s),a,(p',q'))=\delta_{H_{i+1}}((p_s,q_s),a,(p',q'))\wedge\theta(sa)$.
Here $\theta(sa)=\wedge\{\delta_{H_{i+1}}((p',q'),\mu,(p_\mu,q_\mu)):
0<\delta_{H_{i+1}}((p',q'),\mu,(p_\mu,q_\mu))<\delta(q',\mu,q_\mu)\mbox{ and }\mu\in E^*_{uc}\}$.
Note that by the definition of extended fuzzy transition function, the set of distinct membership
grades is finite. Thereby, the process is guaranteed to converge in finite steps. Repeat for all
controllable events appearing in the transitions of $H_{i+1}$, and call the resulting automaton
$H'$.

Step 5): $\LL(H')=\K^\uparrow$.

We give a simple example that illustrates the above algorithm.
\begin{Eg}Let $E=\{a,b\}$ and $E_{uc}=\{b\}$. The automata $G$ and $H$ that generate $\LL$ and $\K$ are
depicted in Figure 1, respectively. Observe that $\delta(3,b,2)>0$, whereas
$\delta_{H_0}((D,3),b,(C,2))=0$. So we remove $(D,3)$ and the associated transition from $H_0$,
namely $H\times G$, and find that no more states need to be removed. Then we reassign the
membership grade of $a$ according to Step 4), and thus get $H'$.
\begin{figure}\centering
\includegraphics{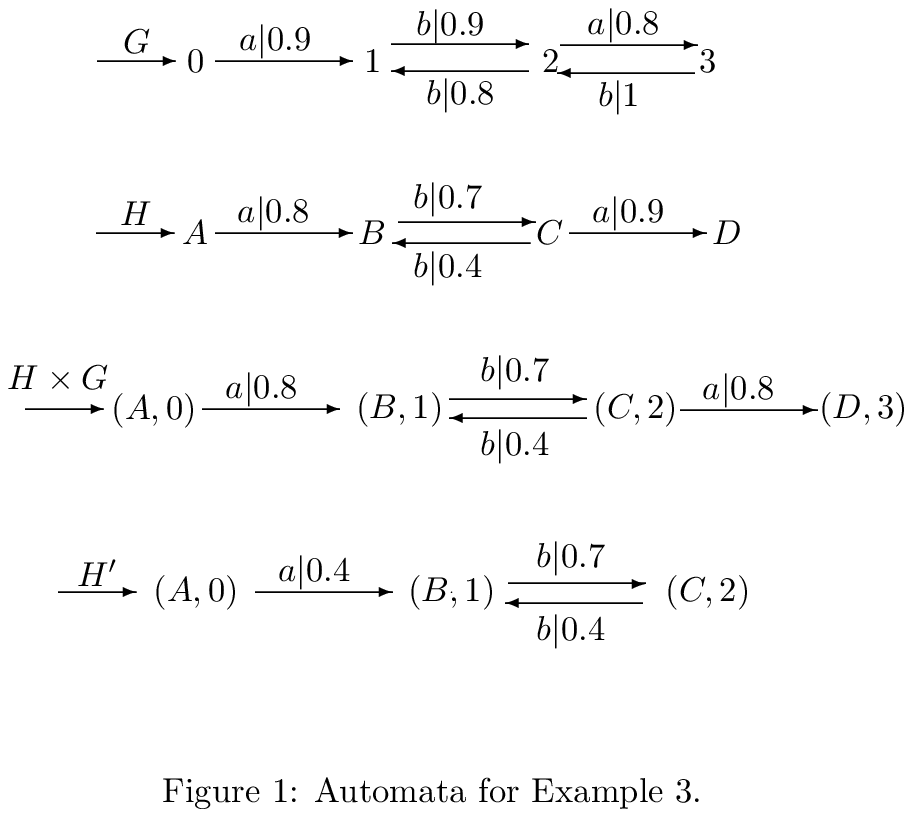}
\end{figure}
\end{Eg}

The following theorem verifies the correctness of the above algorithm.
\begin{Thm}
Let $\LL,\K,G,H_i,H'$ be as in Algorithm for $\K^\uparrow$. Then $\LL(H')=\K^\uparrow$.
\end{Thm}
\begin{proof}
By Theorem \ref{Tuparrow}, it suffices to prove that $\LL(H')=\hat{\K}$, i.e.,
$\LL(H')(s)=\hat{\K}(s)$ for all $s\in E^*$. We prove it by induction on the length of $s$.

For $s=\epsilon$, we know by definition that $\hat{\K}(\epsilon)=0$ if and only if there exists
$\omega\in E^*_{uc}$ such that $\K(\omega)<\LL(\omega)$. If one of these $\omega$ such that
$\LL(H_i)(\omega)=0$, then Steps 2) and 3) force that $H'$ is the empty automaton; otherwise, these
$\omega$ survive in $H_{i+1}$, then by Step 4) we have that $H'$ is also the empty automaton. Hence
$\LL(H')(\epsilon)=0$. Conversely, if $\LL(H')(\epsilon)=0$, then it follows immediately that $H'$
is the empty automaton. By the above algorithm, we know that there must be $\omega\in E^*_{uc}$
such that $\K(\omega)<\LL(\omega)$. Consequently, $\hat{\K}(\epsilon)=0$, which establishes the
basis step.

For the induction step, suppose that $\LL(H')(s)=\hat{\K}(s)$ for all $s\in E^*$ with $|s|\leq n$.
Let us consider $sa$. Note that fuzzy automata under consideration are deterministic finite. Thus
if $a\in E_{uc}$, then by construction and induction hypothesis we obtain that
\begin{eqnarray*}
\LL(H')(sa)&=&\LL(H')(s)\wedge\delta_{H'}((p_s,q_s),a,(p',q'))\\
 &=&\hat{\K}(s)\wedge\delta_{H'}((p_s,q_s),a,(p',q'))\\
 &=&\hat{\K}(s)\wedge\delta_{H_0}((p_s,q_s),a,(p',q'))
 \\&=&\hat{\K}(s)\wedge\K(s)\wedge\delta_{H_0}((p_s,q_s),a,(p',q'))\\
 &=&\hat{\K}(s)\wedge\K(sa)\\&=&\hat{\K}(sa),
 \end{eqnarray*} that is, $\LL(H')(sa)=\hat{\K}(sa)$. If $a\in E_c$, then by definitions we have
 that
\begin{eqnarray*}
\LL(H')(sa)&=&\LL(H')(s)\wedge\delta_{H'}((p_s,q_s),a,(p',q'))\\
 &=&\hat{\K}(s)\wedge\delta_{H_{i+1}}((p_s,q_s),a,(p',q'))\wedge\theta(sa)\\
 &=&\hat{\K}(s)\wedge\delta_{H_{0}}((p_s,q_s),a,(p',q'))\wedge\theta(sa)
 \\&=&\hat{\K}(s)\wedge\K(s)\wedge\delta_{H_{0}}((p_s,q_s),a,(p',q'))\\& &\wedge\theta(sa)
 \\&=&\hat{\K}(s)\wedge\K(sa)\wedge\theta(sa),
 \end{eqnarray*}
and $ \hat{\K}(sa)=\hat{\K}(s)\wedge\K(sa)\wedge\lambda(sa)$. By construction and the definitions
of $\theta(sa)$ and
 $\lambda(sa)$, it is easy to show that $\K(sa)\wedge\theta(sa)=\K(sa)\wedge\lambda(sa)$. Thus we also get that
  $\LL(H')(sa)=\hat{\K}(sa)$ when $a\in E_{c}$. This completes the proof.
\end{proof}

\section{Infimal Controllable Fuzzy Superlanguage}

In parallel with the last section, we now study the ``least" controllable fuzzy superlanguage of
$\K$ mentioned earlier and its calculation when $\K$ is not controllable.

Let us define $\K^\downarrow=\bigcap\limits_{\M\in\mathscr{C}_{sup}(\K)}\M$. Similar to Proposition
\ref{PLargCon}, we have the following result which shows us the existence of the least controllable
fuzzy superlanguage of $\K$.

\begin{Prop}The fuzzy language $\K^\downarrow$ is
the least controllable fuzzy superlanguage of $\K$.\label{PLeastCon}
\end{Prop}
\begin{proof}It follows directly from Proposition \ref{PContr}.
\end{proof}

Since by definition $\K^\downarrow\subseteq\M$ for any $\M\in\mathscr{C}_{sup}(\K)$, we call
$\K^\downarrow$ the {\it infimal controllable fuzzy superlanguage} of $\K$. If $\K$ is
controllable, then $\K^\downarrow=\K$. In the ``worst" case, $\K^\downarrow=\LL$.

Let us proceed to find more characterizations of $\K^\downarrow$.

In order to give a compact formula for $\K^\downarrow$, we introduce a new fuzzy subset $\E^*_{uc}$
of $E^*$, which is given by

\begin{displaymath}
\E^*_{uc}(\omega)=\left\{ \begin{array}{ll}
1, & \textrm{if $\omega\in E^*_{uc}$}\\
0, & \textrm{otherwise}.
\end{array} \right.
\end{displaymath}

We make the following observation on $\E^*_{uc}$.
\begin{Lem}
$(\K\E^*_{uc})(sa)=(\K\E^*_{uc})(s)$ for any $s\in E^*$ and $a\in E_{uc}$.\label{Lsa=s}
\end{Lem}
\begin{proof}By definition, we know that $(\K\E^*_{uc})(s)=\vee_{s_1s_2=s}(\K(s_1)\wedge\E^*_{uc}(s_2))$ and
$(\K\E^*_{uc})(sa)=\vee_{t_1t_2=sa}(\K(t_1)\wedge\E^*_{uc}(t_2))$. Since $|s|$ is finite, we can
assume that $\vee_{s_1s_2=s}(\K(s_1)\wedge\E^*_{uc}(s_2))=\K(s'_1)\wedge\E^*_{uc}(s'_2)$ for some
$s'_1$ and $s'_2$ satisfying $s'_1s'_2=s$. Setting $t'_1=s'_1$ and $t'_2=s'_2a$, we have that
$t'_1t'_2=sa$ and $\K(t'_1)\wedge\E^*_{uc}(t'_2)=\K(s'_1)\wedge\E^*_{uc}(s'_2)$. As a result,
$(\K\E^*_{uc})(sa)=\vee_{t_1t_2=sa}(\K(t_1)\wedge\E^*_{uc}(t_2))\geq\K(t'_1)\wedge\E^*_{uc}(t'_2)
=\K(s'_1)\wedge\E^*_{uc}(s'_2)=(\K\E^*_{uc})(s)$, that is, $(\K\E^*_{uc})(sa)\geq(\K\E^*_{uc})(s)$.
On the other hand, we observe that $\K\E^*_{uc}\in\F\LL$ from Proposition \ref{PFLan}, so
$(\K\E^*_{uc})(sa)\leq(\K\E^*_{uc})(s)$, thus finishing the proof.
\end{proof}

For later need, we define inductively another fuzzy subset $\check{\K}$ of $E^*$ as follows:

$\qquad\quad\ \check{\K}(\epsilon)=1$;

\begin{displaymath}
\check{\K}(sa)=\left\{ \begin{array}{ll}
\K(sa), & \textrm{if $a\in E_{c}$}\\
\check{\K}(s)\wedge\LL(sa), & \textrm{if $a\in E_{uc}$}
\end{array} \right.
\end{displaymath}for all $s\in E^*$ and $a\in E$.

We first present a compact formula for $\check{\K}$.
\begin{Prop}
$\check{\K}=\K\E^*_{uc}\cap\LL$; furthermore, $\check{\K}\in\F\LL$.\label{PKFuzzLan}
\end{Prop}
\begin{proof}We prove the first part by induction on the length of strings $s$. In the case
of $s=\epsilon$, we see that $\check{\K}(\epsilon)=(\K\E^*_{uc}\cap\LL)(\epsilon)=1$, so the basis
step holds. Assume that $\check{\K}(s)=(\K\E^*_{uc}\cap\LL)(s)$ for all strings $s$ with $|s|\leq
n$. Now consider the case of $sa$, where $a\in E$. In the subcase of $a\in E_c$, we know that
$\check{\K}(sa)=\K(sa)$ by definition. On the other hand,
$(\K\E^*_{uc}\cap\LL)(sa)=\K(sa)\wedge\LL(sa)=\K(sa)$. Consequently,
$\check{\K}(sa)=(\K\E^*_{uc}\cap\LL)(sa)$ when $a\in E_c$. In the other subcase $a\in E_{uc}$, we
have that
\begin{eqnarray*}
\check{\K}(sa)&=&\check{\K}(s)\wedge\LL(sa)\mbox{ (by definition)}\\
 &=&(\K\E^*_{uc}\cap\LL)(s)\wedge\LL(sa)\mbox{ (by induction hypothesis)} \\
 &=&(\K\E^*_{uc})(s)\wedge\LL(s)\wedge\LL(sa)
 \\&=&(\K\E^*_{uc})(s)\wedge\LL(sa),
 \end{eqnarray*}and
\begin{eqnarray*}
(\K\E^*_{uc}\cap\LL)(sa)
 &=&(\K\E^*_{uc})(sa)\wedge\LL(sa)\\
 &=&(\K\E^*_{uc})(s)\wedge\LL(sa)\mbox{ (by Lemma \ref{Lsa=s})}.
\end{eqnarray*}
Hence we also have that $\check{\K}(sa)=(\K\E^*_{uc}\cap\LL)(sa)$ when $a\in E_{uc}$. This
completes the proof of the first part.

The second part follows from the first part and Proposition \ref{PFLan}, since $\K$, $\E^*_{uc},$
and $\LL$ are fuzzy languages.
\end{proof}

Our intention of defining $\check{\K}$ is to represent $\K^\downarrow$; we establish the following
theorem.
\begin{Thm}
$\K^\downarrow=\check{\K}.$
\end{Thm}
\begin{proof}
We first show that $\K^\downarrow\subseteq\check{\K}$. It is sufficient to verify that
$\check{\K}\in\mathscr{C}_{sup}(\K)=\{\M\in\F\LL:\K\subseteq\M\subseteq\LL\mbox{ and $\M$ is
controllable}\}$. By Proposition \ref{PKFuzzLan}, we have that $\check{\K}\in\F\LL$ and
$\check{\K}\subseteq\LL$. In order to show that $\K\subseteq\check{\K}$, we prove that
$\K(s)\leq\check{\K}(s)$ by using induction on the length of strings $s$. Obviously, the case
$s=\epsilon$ holds. Assuming that $\K(s)\leq\check{\K}(s)$ for all strings $s$ with $|s|\leq n$, we
next show that $\K(sa)\leq\check{\K}(sa)$ for any $a\in E$. If $a\in E_c$, then
$\K(sa)=\check{\K}(sa)$ by definition. If $a\in E_{uc}$, then we obtain that
\begin{eqnarray*}
\K(sa)
 &\leq&\K(s)\wedge\LL(sa)\\
 &\leq&\check{\K}(s)\wedge\LL(sa)\mbox{ (by assumption)}\\
 &=&\check{\K}(sa)\mbox{ (by definition)}.
\end{eqnarray*}Therefore the induction step holds, too.
It remains to prove that $\check{\K}$ is controllable. In fact, for any $s\in E^*$ and $a\in
E_{uc}$, we always have that $\check{\K}(s)\wedge\LL(sa)=\check{\K}(sa)$ by the definition of
$\check{\K}$. Hence, $\check{\K}$ is controllable by Proposition \ref{Puc}, and thus
$\check{\K}\in\mathscr{C}_{sup}(\K)$.

Next, we prove the other direction of the inclusion, namely, $\check{\K}\subseteq\K^\downarrow$.

Taking any $\M\in\mathscr{C}_{sup}(\K)$, we claim that $\K\E^m_{uc}\cap\LL\subseteq\M$ for any
nonnegative integer $m$, where $\E^m_{uc}$ stands for the concatenation of $m$ copies of $\E_{uc}$.
We prove it by induction on $m$. If $m=0$, it is obvious that $\K\cap\LL\subseteq\M$ since
$\K\subseteq\M$. Suppose that the claim holds for $m$. Let us consider the case $m+1$. Clearly, we
see that $(\K\E^{m+1}_{uc}\cap\LL)(\epsilon)=0\leq1=\M(\epsilon)$, and
$(\K\E^{m+1}_{uc}\cap\LL)(sa)=0\leq\M(sa)$ for any $s\in E^*$ and $a\in E_c$. If $a\in E_{uc}$,
then we have that
\begin{eqnarray*}
(\K\E^{m+1}_{uc}\cap\LL)(sa)
 &=&(\K\E^{m}_{uc})(s)\wedge\LL(sa)\\
 &=&(\K\E^{m}_{uc})(s)\wedge\LL(s)\wedge\LL(sa)\\&\ &\mbox{\qquad\qquad(since $\LL(s)\geq\LL(sa)$)}
 \\&=&(\K\E^{m}_{uc}\cap\LL)(s)\wedge\LL(sa)
 \\&\leq&\M(s)\wedge\LL(sa)\\&\ &\mbox{\qquad\qquad(by induction hypothesis)}
 \\&=&\M(sa)\mbox{ (by the controllability of $\M$)},
 \end{eqnarray*}i.e., $(\K\E^{m+1}_{uc}\cap\LL)(sa)\leq\M(sa)$. So the case of $m+1$ holds,
  finishing the proof of the claim.

 For any $s\in E^*$, we now prove that
 $\check{\K}(s)\leq\M(s)$. By Proposition \ref{PKFuzzLan}, we get that
\begin{eqnarray*}
\check{\K}(s)
 &=&(\K\E^*_{uc})(s)\wedge\LL(s) \\
 &=&[\vee_{s_1s_2=s}(\K(s_1)\wedge\E^*_{uc}(s_2))]\wedge\LL(s)=:(5).
 \end{eqnarray*}Since $|s|$ is finite, there exist $s'_1$ and $s'_2$ satisfying $s'_1s'_2=s$ such that
 $\vee_{s_1s_2=s}(\K(s_1)\wedge\E^*_{uc}(s_2))=\K(s'_1)\wedge\E^*_{uc}(s'_2)$. Hence,
\begin{eqnarray*}
\mbox{(5)}&=&\K(s'_1)\wedge\E^*_{uc}(s'_2)\wedge\LL(s)
 \\&=&\K(s'_1)\wedge\E^{|s'_2|}_{uc}(s'_2)\wedge\LL(s)\\
&\leq&(\K\E^{|s'_2|}_{uc})(s)\wedge\LL(s) \\
&=&(\K\E^{|s'_2|}_{uc}\cap\LL)(s) \\
&\leq&\M(s).
 \end{eqnarray*}The last inequality follows from the foregoing claim that
 $\K\E^m_{uc}\cap\LL\subseteq\M$. Thereby $\check{\K}\subseteq\M$. Furthermore, since $\M$ was
 arbitrary, we obtain that
 $\check{\K}\subseteq\bigcap\limits_{\M\in\mathscr{C}_{sup}(\K)}\M=\K^\downarrow$. Thus the proof of
the theorem is completed.
\end{proof}

Summarily, we have obtained the following characterizations of the least controllable fuzzy
superlanguage $\K^\downarrow$ of $\K$.
\begin{Thm}$\K^\downarrow=\bigcap\limits_{\M\in\mathscr{C}_{sup}(\K)}\M=\K\E^*_{uc}\cap\LL$. More
explicitly, $\K^\downarrow$ can be defined inductively as follows:

$\qquad\quad\ \K^\downarrow(\epsilon)=1$;

\begin{displaymath}
\K^\downarrow(sa)=\left\{ \begin{array}{ll}
\K(sa), & \textrm{if $a\in E_{c}$}\\
\K^\downarrow(s)\wedge\LL(sa), & \textrm{if $a\in E_{uc}$}
\end{array} \right.
\end{displaymath}for all $s\in E^*$ and $a\in E$.\label{TK_}
\end{Thm}

Using the above theorem, we can obtain the support of $\K^\downarrow$ from those of $\K$ and $\LL$.
\begin{Coro}
$supp(\K^\downarrow)=supp(\K)E^*_{uc}\cap supp(\LL)$.
\end{Coro}
\begin{proof}
Note that $supp(\E^*_{uc})=E^*_{uc}$ by definition. The proof follows directly from Theorem
\ref{TK_} and Proposition \ref{Psupp}.
\end{proof}

Based on Theorem \ref{TK_}, we now present an algorithm for the computation of $\K^\downarrow$ from
automata $H$ and $G$ that generate $\K$ and $\LL$, respectively.

Let $H=(Q_H,E,\delta_H,q_{0H})$ and $G=(Q,E,\delta,q_0)$.

{\it Algorithm for $\K^\downarrow$:}

Step 1). Build a fuzzy automaton that generates $\K\E_{uc}^*$ as follows. Add a new state $q_a$ to
$Q_H$. For all $q\in Q_H\cup\{q_a\}$ and all $a\in E_{uc}$, add an arc with label $a|1$ from $q$ to
$q_a$. Call the resulting automaton $H_1$. More formally, $H_1=(Q_1,E,\delta_1,q_{01})$ is given by

$Q_1=Q_H\cup\{q_a\}$, $q_{01}=q_{0}$, and
\begin{displaymath}
\delta_1(p,a,q)=\left\{ \begin{array}{ll}
\delta_H(p,a,q), & \textrm{if $p,q\in Q_H$}\\
1, & \textrm{if $a\in E_{uc}$ and $q=q_a$}\\
0, & \textrm{otherwise.}
\end{array} \right.
\end{displaymath}

Step 2). Take the product operation $H_1\times G$. Then it follows immediately that $\LL(H_1\times
G)=\K^\downarrow$.

The correctness of this algorithm is easily verified once one observes that Step 1) serves to the
concatenation and the Kleene closure in Theorem \ref{TK_}.

The following example illustrates the algorithm.

\begin{Eg}Let $E=\{a,b\}$ and $E_{uc}=\{b\}$. The automata $G$ and $H$ that generate $\LL$ and $\K$ are
depicted in Figure 2, respectively. By adding a new state $N$ and associated transitions in $H$, we
get $H_1$; further, we have the automaton $H_1\times G$ that generates $\K^\downarrow$.
\begin{figure}\centering
\includegraphics{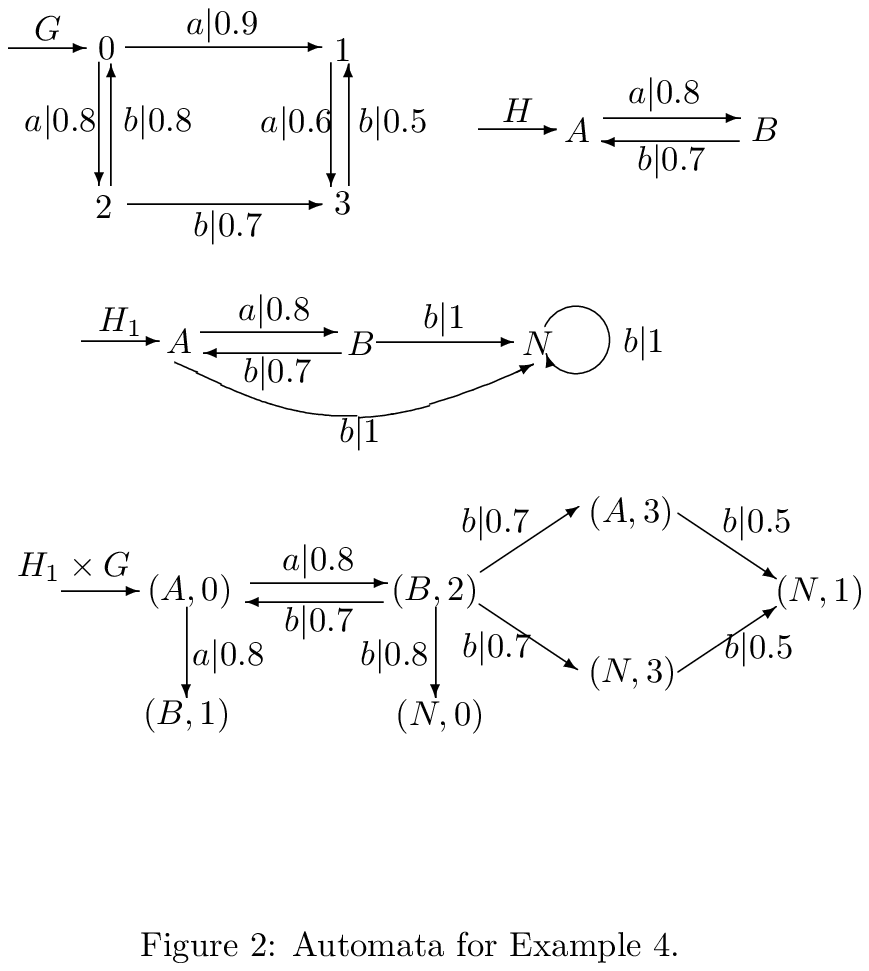}
\end{figure}
\end{Eg}

Now, with the help of Theorem \ref{TK_} we can prove the following fact.
\begin{Prop}Let $\K_1,\K_2\in\F\LL$.

1) \ If $\K_1\subseteq\K_2$, then $\K_1^\downarrow\subseteq\K_2^\downarrow$.

2) \ $(\K_1\cap\K_2)^\downarrow=\K_1^\downarrow\cap\K_2^\downarrow$.

3) \ $(\K_1\cup\K_2)^\downarrow=\K_1^\downarrow\cup\K_2^\downarrow$.
\end{Prop}
\begin{proof}
The assertion 1) is clear by the definition of $\K^\downarrow$. The assertions 2) and 3) can be
proved in the same way using induction on the length of strings, so we only prove 2). Using Theorem
\ref{TK_}, we have that
$(\K_1\cap\K_2)^\downarrow(\epsilon)=1=(\K_1^\downarrow\cap\K_2^\downarrow)(\epsilon)$. Suppose
that 2) holds for all strings $s$ with $|s|\leq n$, namely,
$(\K_1\cap\K_2)^\downarrow(s)=(\K_1^\downarrow\cap\K_2^\downarrow)(s)$. We now consider the case of
$sa$. If $a\in E_c$, then by Theorem \ref{TK_} we get that
$(\K_1\cap\K_2)^\downarrow(sa)=(\K_1\cap\K_2)(sa)=\K_1(sa)\wedge\K_2(sa)=\K_1^\downarrow(sa)\wedge\K_2^\downarrow(sa)
=(\K_1^\downarrow\cap\K_2^\downarrow)(sa).$ If $a\in E_{uc}$, then by induction hypothesis and
Theorem \ref{TK_} we have that
\begin{eqnarray*}
(\K_1\cap\K_2)^\downarrow(sa)
 &=&(\K_1\cap\K_2)^\downarrow(s)\wedge\LL(sa)\\
 &=&(\K_1^\downarrow\cap\K_2^\downarrow)(s)\wedge\LL(sa)
 \\&=&\K_1^\downarrow(s)\wedge\K_2^\downarrow(s)\wedge\LL(sa)
 \\&=&(\K_1^\downarrow(s)\wedge\LL(sa))\wedge(\K_2^\downarrow(s)\wedge\LL(sa))
 \\&=&\K_1^\downarrow(sa)\wedge\K_2^\downarrow(sa)\\&=&(\K_1^\downarrow\cap\K_2^\downarrow)(sa),
 \end{eqnarray*}
finishing the proof of 2).
\end{proof}

Recall that SCP mentioned in Section IV requires us to find a fuzzy supervisor $S$ for the fuzzy
DES $G$ such that the behavior of controlled system lies between the minimal acceptable behavior
$\LL_a$ and the maximal legal behavior $\LL_l$. We end this section with the solution to SCP.
\begin{Thm}
The following statements are equivalent:

1) \ SCP is solvable;

2) \ $\LL_a\subseteq\LL_l^\uparrow$;

3) \ $\LL_a^\downarrow\subseteq\LL_l$.
\end{Thm}
\begin{proof}We only prove that 1) and 2) are equivalent. The equivalence of 1) and 3) can be
proved similarly.

1)$\Rightarrow$2). Since SCP is solvable, there exists a fuzzy supervisor $S$ such that
$\LL_a\subseteq\LL^S\subseteq\LL_l$. By Theorem \ref{Tcont}, we see that $\LL^S$ is controllable.
So $\LL^{S\uparrow}=\LL^S\subseteq\LL_l$; furthermore, $\LL^{S\uparrow}\subseteq\LL_l^\uparrow$.
Consequently, $\LL_a\subseteq\LL_l^\uparrow$.

2)$\Rightarrow$1). Obviously, $\LL_l^\uparrow$ is nonempty and controllable. Thus by Theorem
\ref{Tcont} there exists a fuzzy supervisor $S$ such that $\LL^S=\LL_l^\uparrow$. From the
condition $\LL_a\subseteq\LL_l^\uparrow$ and the fact $\LL_l^\uparrow\subseteq\LL_l$, we obtain
that $\LL_a\subseteq\LL^S\subseteq\LL_l$. Hence, SCP is solvable.
\end{proof}

\section{Conclusion}In this paper, we have formalized the supervisory control theory for fuzzy
DES which have been modelled by (maxmin) fuzzy automata, by extending the prior supervisory control
theory for crisp DES in a natural way. The behavior of fuzzy DES has been described by fuzzy
languages; the supervisors are event feedback and fuzzy. The control objective in this setting is
to find a fuzzy supervisor such that the controlled system only generates legal strings which must
occur with certain minimum membership grades. Our framework generalizes that of Ramadge-Wonham and
reduces to Ramadge-Wonham framework when membership grades in all fuzzy languages must be either
$0$ or $1$.

In the present framework, we have focused on controlled behavior that can be achieved when
controlling a fuzzy DES $G$ with a full-observation fuzzy supervisor $S$. Further research on
supervisory control under partial observation is the subject of work in progress. In addition, if
we restrict the behavior of a fuzzy DES $G$ to a non-fuzzy ``threshold language":
$L(\theta)=\{\omega:\omega\in E^*$ and $\LL(G)(\omega)\geq\theta\}$, where $0<\theta\leq1$, the
corresponding supervisory control may be of interest. A potential application of the supervisory
control theory for fuzzy DES in this paper goes to the design and analysis of dynamic control
systems whose behavior or control rules are acquired from experience of human experts and described
in natural languages.

\end{document}